\documentclass[12pt]{article}
\usepackage{mathrsfs}
\usepackage{graphicx}
\usepackage{amsmath}
\usepackage{amssymb}
\usepackage{caption2}
\begin{document}
\newcommand  {\ba} {\begin{eqnarray}}
\newcommand  {\be} {\begin{equation}}
\newcommand  {\ea} {\end{eqnarray}}
\newcommand  {\ee} {\end{equation}}
\renewcommand{\thefootnote}{\fnsymbol{footnote}}
\renewcommand{\figurename}{Figure.}
\renewcommand{\captionlabeldelim}{.~}

\vspace*{1cm}
\begin{center}
 {\Large\textbf{The Matter-Antimatter Asymmetry and Cold Dark Matter from The Left-Right Mirror Symmetric Model with The Global $U(1)_{B-L}\otimes U(1)_{D}$}}

\vspace{1cm}
 \textbf{Wei-Min Yang}

\vspace{0.4cm}
 \emph{Department of Modern Physics, University of Science and Technology of China, Hefei 230026, P. R. China}

\vspace{0.2cm}
 \emph{E-mail: wmyang@ustc.edu.cn}
\end{center}

\vspace{1cm}
\noindent\textbf{Abstract}: The paper suggests a left-right mirror symmetric model with the global $U(1)_{B-L}\otimes U(1)_{D}$ symmetries. The model can simultaneously accommodate the standard model, neutrino physics, matter-antimatter asymmetry and cold dark matter. The model naturally and elegantly accounts for the origin of the tiny neutrino mass, matter-antimatter asymmetry and cold dark matter. In particular, it predicts a number of interesting results, e.g. a right-handed neutrino asymmetry and a dark Goldstone boson. It is also feasible and promising to test the model in future experiments.

\vspace{1cm}

\vspace{0.3cm}

\newpage
 \noindent\textbf{I. Introduction}

\vspace{0.3cm}
 The standard model (SM) has been evidenced to be a very successful theory at the electroweak energy scale. The precise tests for the SM physics have established plenty of knowledge about the elementary particles \cite{1,2}. Nevertheless, there are a number of the unsolved issues in the particle physics and universe observations, which are not able to be accounted by the SM \cite{3,4}. The issues in flavor physics are the two facts, i) the mass spectrum hierarchy of the quarks and charged leptons \cite{5}, ii) the distinct difference between the quark flavor mixing pattern and the lepton one \cite{6}. The issues in neutrino physics include what is the exact cause of the Sub-eV neutrino masses \cite{7}? Is the neutrino nature Dirac or Majorana fermion? In other words, is there $0\nu\beta\beta$ or not \cite{8}? Is the $CP$ violation in the lepton mixing vanishing or not? The issues in the universe are more difficult. What is really mechanism of the genesis of the matter-antimatter asymmetry \cite{9}? Whether it has to do with the $CP$ violation in the quark and/or lepton sector or not \cite{10}? What is the nature of the cold dark matter \cite{11}? And so on. All of the problems are very important and significant for both particle physics and cosmology, so they attract great attentions in the experiment and theory fields all the time \cite{12}.

 The researches for the above-mentioned problems have motivated many new theories beyond the SM. The various theoretical suggestions have been proposed to solve them \cite{13}. Some grand unification models based on the $SO(10)$ gauge group can give some reasonable interpretations for fermion masses and flavor mixings \cite{14a}. Some kinds of flavor family symmetry are employed to understand the neutrino mixing \cite{14a}. The references \cite{15} discusses a combination of GUT and flavor symmetry. The neutrino mass can be implemented by the see-saw mechanism \cite{16}. The baryon asymmetry can be achieved through the electroweak baryogenesis or leptogenesis \cite{17a}, and the references \cite{17b} also gave the Dirac leptogenesis idea. The cold dark matter candidate can be scalar boson dark matter \cite{18}, sterile neutrino dark matter \cite{19a}, supersymmetry dark matter\cite{19b}, and so on \cite{20a}. The references \cite{20b} studied the mirror symmetric models. These theories are successful in explaining one or two specific aspects of the problems, but it seems very difficult for them to solve many aspects of the problems simultaneously. Indeed, it is especially hard for a model construction to keep the principle of the simplicity, feasibility and the fewer number of parameters, otherwise, the theory will be excessive complexity and incredible. On the basis of the unity of nature, a realistic theory beyond the SM should simultaneously accommodate and account for the neutrino physics, baryon asymmetry and cold dark matter besides the SM, in other words, it has to integrate the four things completely. Therefore, it is still a large challenge for theoretical particle physicists to realize the purpose \cite{21}.

 In this work, I try to construct a simple and feasible particle model. It can simultaneously accommodate the SM, neutrino physics, matter-antimatter asymmetry and cold dark matter. The model local gauge groups are $SU(3)_{C}\otimes SU(2)_{L}\otimes SU(2)_{R}\otimes U(1)_{Y}$. In addition, it has the global $U(1)_{B-L}\otimes U(1)_{D}$ symmetries where $D$ is the dark matter particle number, and a discrete $Z_{2}$ mirror symmetry. The left-right mirror symmetry is a main characteristic of the model. Besides the SM particles, the model includes some new particles as follows, three gauge bosons related to the right-handed isospin subgroup, three scalar fields which are respectively a singlet, a right-handed doublet and a left-right doublet, a singlet Majorana neutrino, and vector-like quark and charged lepton. All kinds of the model symmetries are spontaneously broken step by step by the scalar fields developing hierarchical vacuum expectation values (VEVs) except the global $U(1)_{B-L}$ symmetry. In the model, the left-right doublet scalar field merely develops a tiny VEV due to its very heavy mass by nature, the neutrino tiny masses exactly arise from it. In addition, the left-right doublet scalar boson can decay into a left-handed doublet lepton and an antiparticle state of the right-handed doublet lepton, moreover, the decay is an out-of-equilibrium and $CP$-violating process. The $CP$-violating source lies in the scalar potential, which brings about the mirror asymmetry and $CP$ violation of the scalar potential. The decay process will eventually lead to both the baryogenesis and the right-handed neutrino asymmetry through the mirror symmetry breaking and the electroweak sphaleron process \cite{22}. The singlet Majorana neutrino has some unique natures in the model, for instance, it has the dark matter number instead of the lepton number, and it has only a coupling to the singlet scalar. These properties ensure that it will become a cold dark matter particle. The model can not only completely accommodate all the current experimental data of the SM and neutrino physics, but also correctly reproduce the observed value of the baryon asymmetry and the relic abundance of the cold dark matter. In particular, the model predicts some interesting results, for example, the light right-handed neutrino asymmetry is the same size as the baryon asymmetry, its relic abundance is the same size as one of the microwave background photon, and a dark Goldstone boson exists, and so on. Finally, the model is feasible and promising to be tested in future experiments. I give some methods of searching some of the new particles in the experiments at the LHC \cite{23}.

 The remainder of this paper is organized as follows. In Section II I outline the model. Sec. III introduces the model symmetry breakings and the particle masses and mixings. Sec. IV and Sec. V respectively discuss the matter-antimatter asymmetry and the cold dark matter. Sec. VI is the numerical results and a simple discussion for the experimental search. Sec. VII is devoted to conclusions.

\vspace{1cm}
 \noindent\textbf{II. Model}

\vspace{0.3cm}
 The local gauge symmetries of the model are the direct product groups of $SU(3)_{C}\otimes SU(2)_{L}\otimes SU(2)_{R}\otimes U(1)_{Y}$. The gauge groups are obviously left-right symmetrical. The model particle contents and their gauge quantum numbers are in detail listed as follows,
\begin{alignat}{1}
 & G_{\mu}^{a}(8,1,1,0)\,,\hspace{0.5cm} W_{L\mu}^{i}(1,3,1,0)\,,\hspace{0.5cm} W_{R\mu}^{i}(1,1,3,0)\,,\hspace{0.5cm}
   Y_{\mu}(1,1,1,1)\,,\nonumber\\
 & Q_{L}(3,2,1,\frac{1}{3})\,,\hspace{0.5cm} Q_{R}(3,1,2,\frac{1}{3})\,,\hspace{0.5cm}
   L_{L}(1,2,1,-1)\,,\hspace{0.5cm} L_{R}(1,1,2,-1)\,,\nonumber\\
 & \widetilde{u}_{L,R}(3,1,1,\frac{4}{3})\,,\hspace{0.4cm} \widetilde{d}_{L,R}(3,1,1,-\frac{2}{3})\,,\hspace{0.4cm}
   \widetilde{e}_{L,R}(1,1,1,-2)\,,\hspace{0.4cm} \widetilde{\nu}_{L,R}(1,1,1,0)\,,\nonumber\\
 & H_{L}(1,2,1,1)\,,\hspace{0.5cm} H_{R}(1,1,2,1)\,,\hspace{0.5cm}
   H_{LR}(1,2,\overline{2},0)\,,\hspace{0.5cm} \phi(1,1,1,0)\,.
\end{alignat}
 These notations are self-explanatory, and all kinds of the fermions imply three generations as usual. The first row in (1) is gauge fields. The second row are usual quarks and leptons, which include the left-handed and right-handed doublets, in particular, the light right-handed neutrino $\nu_{R}$ is embodied in $L_{R}$. The third row introduces heavy vector-like quarks and charged leptons, and singlet Majorana neutrinos. Since a vector-like fermion is a singlet under the left-handed and right-handed isospin subgroups, its left-handed and left-handed fields have the same quantum number. In addition, I stress that $\widetilde{\nu}$ is a gauge singlet Majorana fermion, so its left-handed and right-handed components are each other $CP$ conjugation by the relation $\widetilde{\nu}_{R}=\widetilde{\nu}_{L}^{c}=C\overline{\widetilde{\nu}_{L}}^{T}$ where $C$ is a charge conjugation matrix. The last row is four types of scalar fields. Only $\phi$ is a complex singlet, the rest are complex multiplets. Their component representations are such as
\ba
 H_{L/R}=\left(\begin{array}{c}H_{L/R}^{+}\\H_{L/R}^{0}\end{array}\right),\hspace{0.5cm}
 H_{LR}=\frac{1}{\sqrt{2}}\left(\begin{array}{cc}H_{LR}^{0*}&H_{LR}^{+}\\-H_{LR}^{-}&H_{LR}^{0}\end{array}\right).
\ea
 Here the relation $H_{LR}=\tau_{2}H_{LR}^{*}\tau_{2}$ is self-evident where $\tau_{2}$ is the second of Pauli matrices. In addition, the notations $H_{L/R}'=i\tau_{2}H_{L/R}^{*}$ are used hereinafter. Each scalar field is responsible for a special symmetry breaking. In brief, the model extends the SM to the left-right mirror symmetrical theory. The new non-SM particles in (1) will play key roles in the new physics beyond the SM, in particular, in the origin of matter-antimatter asymmetry and cold dark matter.

 In addition to the above local gauge symmetries, the model keeps the global symmetries $U(1)_{B-L}\otimes U(1)_{D}$. For all kinds of the fields in (1), their $B$-$L$ and $D$ quantum numbers are listed as follows,
\begin{alignat}{1}
 & [Q,\widetilde{u},\widetilde{d}]\rightarrow(\frac{1}{3},0)\,,\hspace{0.5cm}[L,\widetilde{e}]\rightarrow(-1,0)\,,\nonumber\\
 & \widetilde{\nu}_{L}\rightarrow(0,1)\,,\hspace{0.5cm}\widetilde{\nu}_{R}\rightarrow(0,-1)\,,\hspace{0.5cm}
   \phi\rightarrow(0,2)\,,\nonumber\\
 & \mbox{the others}\rightarrow(0,0)\,.
\end{alignat}
 Obviously, all the particles in (1) have normal $B$-$L$ quantum numbers except the singlet Majorana neutrino $\widetilde{\nu}$ and scalar $\phi$. $\widetilde{\nu}$ and $\phi$ have actually the dark matter numbers instead of the $B$-$L$ quantum numbers. Only the two particles consist in the dark sector, while the rest of the particles fall into the visible sector. In what follows, $U(1)_{B-L}$ is always an exact and unbroken symmetry, but $U(1)_{D}$ will be broken spontaneously. The model characteristics not only guarantee that $\widetilde{\nu}$ will become a cold dark matter particle, but also lead that the origin of the light neutrino masses will completely be different from one of the quark and charged lepton masses, furthermore, the lepton flavor mixing will greatly be different from the quark one.

 Finally, the model has also a discrete symmetry $Z_{2}$, namely the left-right mirror symmetry. It is defined by the field transforms as follows,
\begin{alignat}{1}
 & f_{L}\leftrightarrow f_{R}\,,\hspace{0.5cm} \widetilde{f}_{L}\leftrightarrow \widetilde{f}_{R}\,,\nonumber\\
 & H_{L}\leftrightarrow H_{R}\,,\hspace{0.5cm} H_{LR}\leftrightarrow H_{LR}^{\dagger}\,,\hspace{0.5cm}
   \phi\leftrightarrow \phi^{*}\,,\nonumber\\
 & W_{L\mu}\leftrightarrow W_{R\mu}\,,\hspace{0.5cm} G_{\mu}\leftrightarrow G_{\mu}\,,\hspace{0.5cm}
   Y_{\mu}\leftrightarrow Y_{\mu}\,,
\end{alignat}
 where $H_{LR}\leftrightarrow H_{LR}^{\dagger}$ means that its component transforms are $H_{LR}^{0}\leftrightarrow H_{LR}^{0*},H_{LR}^{\pm}\leftrightarrow -H_{LR}^{\pm}$. For $G_{\mu}$ and $Y_{\mu}$, both its mirror particle and its antiparticle are exactly itself. For $\widetilde{\nu}_{L}$ and $\phi$, its mirror particle is exactly its antiparticle. Later, we will see that the global and discrete symmetries play key roles in the model.

 Under all kinds of the above-mentioned symmetries, the invariant Lagrangian of the model is composed of the following three parts. Firstly, the gauge kinetic energy terms are
\begin{alignat}{1}
 \mathscr{L}_{Gauge}=
 &\:\mathscr{L}_{pure\,gauge}+i\overline{f}\gamma^{\mu}D_{\mu}f+i\overline{\widetilde{f}}\gamma^{\mu}D_{\mu}\widetilde{f}
  +\frac{i}{2}\overline{\widetilde{\nu}}\gamma^{\mu}\partial_{\mu}\widetilde{\nu}+\partial^{\mu}\phi^{*}\,\partial_{\mu}\phi\nonumber\\
 &+(D^{\mu}H_{L})^{\dagger}(D_{\mu}H_{L})+(D^{\mu}H_{R})^{\dagger}(D_{\mu}H_{R})+Tr[(D^{\mu}H_{LR})^{\dagger}(D_{\mu}H_{LR})]\,,
\end{alignat}
 where $f$ denote the usual fermions in (1), and $\widetilde{f}$ are the vector-like quarks and charged leptons. The covariant derivative $D_{\mu}$ is defined by
\ba
 D_{\mu}=\partial_{\mu}+i\left(g_{s}G_{\mu}^{a}\frac{\lambda^{a}}{2}+gW_{L\mu}^{i}\frac{\tau_{L}^{i}}{2}
 +gW_{R\mu}^{i}\frac{\tau_{R}^{i}}{2}+g_{Y}Y_{\mu}\frac{Q_{Y}}{2}\right),
\ea
 where $\lambda^{a}$ and $\tau^{i}$ are respectively Gell-Mann and Pauli matrices, $Q_{Y}$ is the charge operator of $U(1)_{Y}$, and $g_{s},g,g_{Y}$ are three gauge coupling constants. Since both $\widetilde{\nu}$ and $\phi$ are gauge singlets, their kinetic energy terms are simply and directly written out. The gauge symmetry breakings will lead to gauge field masses and mixing through the Higgs mechanism.

 Secondly, the model Yukawa couplings are given by
\begin{alignat}{1}
  \mathscr{L}_{Yukawa}=
 & \:\overline{Q_{L}}H_{L}'Y_{u}\widetilde{u}_{R}+\overline{\widetilde{u}_{L}}Y_{u}^{\dagger}H_{R}^{\prime\dagger}Q_{R}
   -\overline{\widetilde{u}_{L}}M_{\widetilde{u}}\,\widetilde{u}_{R} \nonumber\\
 & +\overline{Q_{L}}H_{L}Y_{d}\widetilde{d}_{R}+\overline{\widetilde{d}_{L}}Y_{d}^{\dagger}H_{R}^{\dagger}Q_{R}
   -\overline{\widetilde{d}_{L}}M_{\widetilde{d}}\,\widetilde{d}_{R} \nonumber\\
 & +\overline{L_{L}}H_{L}Y_{e}\widetilde{e}_{R}+\overline{\widetilde{e}_{L}}Y_{e}^{\dagger}H_{R}^{\dagger}L_{R}
   -\overline{\widetilde{e}_{L}}M_{\widetilde{e}}\,\widetilde{e}_{R} \nonumber\\
 & +\overline{Q_{L}}H_{LR}Y_{Q}Q_{R}+\overline{L_{L}}H_{LR}Y_{L}L_{R}
   +\overline{\widetilde{\nu}_{L}}\phi Y_{\widetilde{\nu}}\,\widetilde{\nu}_{R}+h.c.\,,
\end{alignat}
 in which the last Majorana term can alternatively be written as $\overline{\widetilde{\nu}_{L}}\widetilde{\nu}_{R}=\overline{\widetilde{\nu}_{R}^{c}}\widetilde{\nu}_{R}
 =\widetilde{\nu}_{R}^{T}C\widetilde{\nu}_{R}$. The couplings $Y_{u,d,e}$ are $3\times3$ complex matrices and $Y_{Q,L,\widetilde{\nu}}$ are $3\times3$ Hermitian matrices. $M_{\widetilde{u},\widetilde{d},\widetilde{e}}$ are Hermitian mass matrices of the vector-like quarks and charged leptons. $M_{\widetilde{f}}$ maybe arise from some flavon VEVs which break the flavor family symmetry. However, here they are directly permitted by the model symmetries. In any case, all of $Y_{Q,L,\widetilde{\nu}}$ and $M_{\widetilde{u},\widetilde{d},\widetilde{e}}$ can be taken as real diagonal matrices by choosing the flavor basic. It should be noted that the $U(1)_{B-L}\otimes U(1)_{D}$ symmetries prohibit such terms as $\overline{L_{L}}H_{L}'\widetilde{\nu}_{R},\,\overline{\widetilde{\nu}_{L}}H_{R}^{\prime\dagger}L_{R}$ and $\overline{\widetilde{f}_{L}}\phi\widetilde{f}_{R},\,\widetilde{\nu}_{L}^{T}M\widetilde{\nu}_{L},\,\widetilde{\nu}_{R}^{T}M\widetilde{\nu}_{R}$. Obviously, $\widetilde{\nu}$ and $\phi$ become two exceptional particles in (7) by virtue of their dark matter nature. The Yukawa couplings of (7) will bring about reasonable explanations for the light neutrino masses, matter-antimatter asymmetry and cold dark matter.

 Thirdly, the model scalar potential is written as
\begin{alignat}{1}
  V_{Scalar}=
 & \:\lambda_{\phi}\left(\phi^{*}\phi-\frac{v_{\phi}^{2}}{2}\right)^{2}
   +\lambda_{LR}\left(TrH_{LR}^{\dagger}H_{LR}-\frac{v_{LR}^{2}}{2}+\frac{\Omega}{2\lambda_{LR}v_{LR}^{2}}\right)^{2}\nonumber\\
 & +\lambda_{H}\left(H_{L}^{\dagger}H_{L}-\frac{v_{L}^{2}}{2}+\frac{\Omega}{2\lambda_{H}v_{L}^{2}}\right)^{2}
   +\lambda_{H}\left(H_{R}^{\dagger}H_{R}-\frac{v_{R}^{2}}{2}+\frac{\Omega}{2\lambda_{H}v_{R}^{2}}\right)^{2} \nonumber\\
 & +(\phi^{*}\phi-\frac{v_{\phi}^{2}}{2})\left(\lambda_{1}(TrH_{LR}^{\dagger}H_{LR}-\frac{v_{LR}^{2}}{2})
   +\lambda_{2}(H_{L}^{\dagger}H_{L}-\frac{v_{L}^{2}}{2}+H_{R}^{\dagger}H_{R}-\frac{v_{R}^{2}}{2})\right) \nonumber\\
 & +\lambda_{3}(TrH_{LR}^{\dagger}H_{LR}-\frac{v_{LR}^{2}}{2})\left(H_{L}^{\dagger}H_{L}-\frac{v_{L}^{2}}{2}
   +H_{R}^{\dagger}H_{R}-\frac{v_{R}^{2}}{2}\right) \nonumber\\
 & +\lambda_{4}(H_{L}^{\dagger}H_{L}-\frac{v_{L}^{2}}{2})(H_{R}^{\dagger}H_{R}-\frac{v_{R}^{2}}{2}) \nonumber\\
 & -2\mu_{0}\left(e^{i\delta}H_{L}^{\dagger}H_{LR}H_{R}+e^{-i\delta}H_{R}^{\dagger}H_{LR}^{\dagger}H_{L}\right),
\end{alignat}
 where $\Omega=\mu_{0}v_{L}v_{R}v_{LR}$ for convenience. All kinds of the parameters in (8) are self-explanatory. All the coupling parameters are real numbers except the phase $\delta$. The self-couplings $[\lambda_{\phi},\lambda_{LR},\lambda_{H}]$ should be $\sim0.1$, but the interactive couplings $[\lambda_{1},\cdots,\lambda_{4}]$ should be very weak. $[v_{\phi},v_{LR},v_{L},v_{R}]$ are respectively the VEVs of the corresponding scalar fields, see the following equation (9). $\mu_{0}$ is a positive parameter with mass dimension, and $e^{i\delta}$ is in fact the complex phase factor of $\mu_{0}$ only but it is visibly written out. It should be pointed out that the terms related to $\Omega$ are actually the original mass terms of the corresponding scalar fields before the model symmetry breakings. It can clearly be seen from (8) that $\delta\neq 0$ explicitly leads to the mirror asymmetry and $CP$ violation in the scalar sector. $v_{L}\neq v_{R}\neq 0$ will spontaneously break both the gauge symmetry and the mirror symmetry, in addition, $v_{\phi}\neq 0$ will spontaneously break the $U(1)_{D}$ symmetry. In short, the scalar potential structures are reasonable and regular, in particular, the freedom of the model parameters is greatly reduced due to the mirror symmetry. In comparison with the SM Higgs sector \cite{24}, however, the model scalar sector is more varied and interesting. In conclusion, the above contents form the theoretical framework of the model.

\vspace{1cm}
 \noindent\textbf{III. Symmetry Breakings and Particle Masses and Mixings}

\vspace{0.3cm}
 The model symmetry breakings are implemented by the scalar potential (8). The potential vacuum configurations are strictly derived from the extreme values of (8) according to the mathematical program. The detailed expressions are as follows,
\begin{alignat}{1}
 & \phi \rightarrow \frac{1}{\sqrt{2}}(\phi_{s}+v_{\phi}+i\phi_{g})\,,\hspace{0.5cm}
   H_{L/R}\rightarrow\frac{1}{\sqrt{2}}\left(\begin{array}{c}0\\H_{L/R}^{0}+\,v_{L/R}\end{array}\right),\nonumber\\
 & H_{LR}\rightarrow\frac{1}{\sqrt{2}}\left(\begin{array}{cc}H_{LR}^{0*}+\frac{v_{LR}}{\sqrt{2}}\,e^{i\delta}&H_{LR}^{+}\\
   -H_{LR}^{-}&H_{LR}^{0}+\frac{v_{LR}}{\sqrt{2}}\,e^{-i\delta}\end{array}\right),\nonumber\\
 & \langle \phi\rangle=\frac{v_{\phi}}{\sqrt{2}}\,,\hspace{0.5cm}
   \langle H_{L/R}\rangle=\frac{v_{L/R}}{\sqrt{2}}\left(\begin{array}{c}0\\1\end{array}\right),\hspace{0.5cm}
   \langle H_{LR}\rangle=\frac{v_{LR}}{2}\left(\begin{array}{cc}e^{i\delta}&0\\0&e^{-i\delta}\end{array}\right).
\end{alignat}
 The above VEVs are estimated as $ v_{\phi}\sim 500, v_{L}\sim 250, v_{R}\sim 10^{8}, v_{LR}\sim 10^{-7}$ (all are in GeV as unit). They show a hierarchy of the symmetry breaking scales. For such VEVs arrangement, the conditions of the vacuum stabilization are derived from the conditions of the potential minimum, they are
\ba
  [\lambda_{\phi},\lambda_{H},\mu_{0}]>0\,,\hspace{0.5cm}
  \mbox{the ordered principal minor of}\: \left|\begin{array}{ccc}2\lambda_{\phi}&\lambda_{2}&\lambda_{2}\\\lambda_{2}&2\lambda_{H}&\lambda_{4}\\
  \lambda_{2}&\lambda_{4}&2\lambda_{H}\end{array}\right|>0\,.
\ea
 Of course, all kinds of the requirements are not difficult to be satisfied so long as the parameters are chosen as some suitable values.

 On physics mechanism, the model symmetry breakings are achieved step by step at different energy scales. First of all, the gauge subgroups $SU(2)_{R}\otimes U(1)_{Y}$ are spontaneously broken down $U(1)_{Y'}$ which is namely the hypercharge symmetry of the SM. This is accomplished by the neutral component of the right-handed doublet $H_{R}$ developing the VEV of $v_{R}\sim 10^{8}$ GeV, meanwhile, this also leads to the mirror symmetry breaking. Secondly, $SU(2)_{L}\otimes U(1)_{Y'}\rightarrow U(1)_{em}$, i.e. the electroweak breaking. It is completed by the neutral component of the left-handed doublet $H_{L}$ developing the VEV of $v_{L}\sim 250$ GeV. Lastly, the neutral component of the left-right doublet $H_{LR}$ also develops the VEV of $v_{LR}\sim 10^{-7}$ GeV together with the complex phase factor $e^{i\delta}$. The reason for this is of course from the $\mu_{0}$ term in (8). In fact, the $\frac{\Omega}{2\lambda_{LR}v_{LR}^{2}}$ term in (8) is proportional to the original mass of the $H_{LR}$ field before the symmetry breakings, see the following equation (11). For very heavy $M_{H_{LR}}\sim 10^{10}$ GeV and $\mu_{0}\sim 10^{4}$ GeV, this inevitably leads to a very small $v_{LR}\sim 10^{-7}$ GeV. As a result, the tiny $v_{LR}$ will give rise to the tiny neutrino masses, see the following equation (13). In fact, this is a type-II seesaw \cite{25}. In the dark sector, the $U(1)_{D}$ symmetry breaking is realized by the singlet $\phi$ developing the VEV of $v_{\phi}\sim 500$ GeV. This breaking generates a pseudo scalar Goldstone boson $\phi_{g}$ and the $\widetilde{\nu}$ mass. Finally, it is again emphasized that the $U(1)_{B-L}$ symmetry is maintained and unbroken all the time.

 After the model symmetry breakings are over, all kinds of the particles are generated their masses and mixings according to the standard procedures. In the scalar sector, the boson masses and mixing are such as
\begin{alignat}{1}
 & \mathscr{L}_{Scalar Mass}=-M_{H_{LR}}^{2}(H_{LR}^{0*}H_{LR}^{0}+H_{LR}^{+}H_{LR}^{-})
   -(\phi_{s},H_{L}^{0},H_{R}^{0})\frac{M_{S}^{2}}{2}(\phi_{s},H_{L}^{0},H_{R}^{0})^{T},\nonumber\\
 & M_{H_{LR}}^{2}=\frac{\Omega}{v_{LR}^{2}}=\frac{\mu_{0}v_{L}v_{R}v_{LR}}{v_{LR}^{2}}\,,\hspace{0.5cm}
   M_{\phi_{g}}^{2}=0\,,\nonumber\\
 & M_{S}^{2}=\left(\begin{array}{ccc}2\lambda_{\phi}v_{\phi}^{2}&\lambda_{2}v_{\phi}v_{L}&\lambda_{2}v_{\phi}v_{R}\\
   \lambda_{2}v_{\phi}v_{L}&2\lambda_{H}v_{L}^{2}+\frac{\Omega}{v_{L}^{2}}&\lambda_{4}v_{L}v_{R}-\mu_{0}v_{LR}\\
   \lambda_{2}v_{\phi}v_{R}&\lambda_{4}v_{L}v_{R}-\mu_{0}v_{LR}&2\lambda_{H}v_{R}^{2}+\frac{\Omega}{v_{R}^{2}}\end{array}\right),
\end{alignat}
 where the dark Goldstone $\phi_{g}$ has no any mass and mixing. In fact, the mass of $H_{LR}^{0}$ has also a tiny term $2\lambda_{LR}v_{LR}^{2}$, but it can surely be omitted because of $v_{LR}\ll v_{L}\sim v_{\phi}\ll v_{R}$. By the same token, the mixing between $H_{LR}^{0}$ and the three neutral bosons $\phi_{s},H_{L}^{0},H_{R}^{0}$ can be ignored as well. Therefore, I only consider the mixing of the latter three bosons. Their mass eigenvalues and mixing angles can be obtained by diagonalizing $M_{S}^{2}$. It should be noted that the mass-squared matrix $M_{S}^{2}$ must keep being positive definite in order that its eigenvalues are all positive, therefore, the relevant coupling parameters have to satisfy some restrictions. These restrictions are exactly equivalent to the vacuum stabilization conditions (10) if all the smaller terms involving in $v_{LR}$ are ignored. When $\lambda_{2}$ and $\lambda_{4}$ are smaller, the diagonal elements of $M_{S}^{2}$ approximate to $M_{\phi_{s}}^{2},M_{H_{L}^{0}}^{2},M_{H_{R}^{0}}^{2}$. At the present day, $M_{H_{L}^{0}}$ has been measured as $125$ GeV by the LHC \cite{26}. The model will predict that $M_{\phi_{s}}$ is around several hundred GeVs, $M_{H_{R}^{0}}\sim 10^{7}$ GeV and $M_{H_{LR}}\sim 10^{10}$ GeV. $H_{R}^{0}$ and $H_{LR}$ are too heavy to be detected, but it is possible to find the neutral boson $\phi_{s}$ at the LHC. $\phi_{g}$ is actually a species of the hot dark matter particles, so it inhabits in the dark matter sector.

 In the gauge sector, the gauge symmetry breakings bring about masses and mixings of the vector gauge bosons through the Higgs mechanism. The detailed results are as follows,
\begin{alignat}{1}
 &g(W_{L\mu}^{i}\frac{\tau_{L}^{i}}{2}+W_{R\mu}^{i}\frac{\tau_{R}^{i}}{2})+g_{Y}Y_{\mu}\frac{Q_{Y}}{2}\longrightarrow\nonumber\\
 & \frac{g}{\sqrt{2}}(W_{L\mu}^{+}\tau_{L}^{+}+W_{L\mu}^{-}\tau_{L}^{-}+W_{R\mu}^{+}\tau_{R}^{+}+W_{R\mu}^{-}\tau_{R}^{-})
   +g(Z_{\mu}Q_{L}+\widetilde{Z}_{\mu}Q_{R})+eA_{\mu}Q_{e}\,,\nonumber\\
 & tan\widetilde{\theta}=\frac{g_{Y}}{g}\,,\hspace{0.5cm}
   tan\theta_{W}=sin\widetilde{\theta}\,,\hspace{0.5cm} e=gsin\theta_{W}\,,\nonumber\\
 & Q_{e}=I^{L}_{3}+I^{R}_{3}+\frac{Q_{Y}}{2}\,,\hspace{0.5cm}
   Q_{L}=\frac{I^{L}_{3}-sin^{2}{\theta_{W}}Q_{e}}{cos\theta_{W}}\,,\hspace{0.5cm}
   Q_{R}=\frac{I^{R}_{3}-sin^{2}{\widetilde{\theta}}(Q_{e}-I^{L}_{3})}{cos\widetilde{\theta}}\,,\nonumber\\
 & \left(\begin{array}{c}A_{\mu}\\Z_{\mu}\\\widetilde{Z}_{\mu}\end{array}\right)
   =\left(\begin{array}{ccc}cos\theta&sin\theta&0\\-sin\theta&cos\theta&0\\0&0&1\end{array}\right)
   \left(\begin{array}{ccc}cos\widetilde{\theta}&0&sin\widetilde{\theta}\\0&1&0\\-sin\widetilde{\theta}&0&cos\widetilde{\theta}\end{array}\right)
   \left(\begin{array}{c}Y_{\mu}\\W^{3}_{L\mu}\\W^{3}_{R\mu}\end{array}\right),\nonumber\\
 & M_{W_{L\mu}}=\frac{gv_{L}}{2}\,,\hspace{0.4cm}M_{W_{R\mu}}=\frac{gv_{R}}{2}\,,\hspace{0.4cm}
   M_{Z_{\mu}}=\frac{M_{W_{L\mu}}}{cos\theta_{W}}\,,\hspace{0.4cm}
   M_{\widetilde{Z}_{\mu}}=\frac{M_{W_{R\mu}}}{cos\widetilde{\theta}}\,,\hspace{0.4cm}
   M_{A_{\mu}}=0\,.
\end{alignat}
 In (12), there are only two independent parameters, namely the two gauge coupling constants $g$ and $g_{Y}$. $\widetilde{\theta}$ is a mixing angle for the right-handed isospin symmetry breaking, and $\theta_{W}$ is a mixing angle for the left-handed isospin symmetry breaking. $Q_{L}$ and $Q_{R}$ are two charge operators related to the two massive neutral gauge fields $Z_{\mu}$ and $\widetilde{Z}_{\mu}$, respectively. In addition, it should be pointed out that the mixing angle between $Z_{\mu}$ and $\widetilde{Z}_{\mu}$ is $\sim\frac{v_{L}^{2}cos\widetilde{\theta}\,tan^{2}\theta_{W}}{v_{R}^{2}cos\theta_{W}}$, it is too small so that I can leave it out. Similarly, the mixing angle between $W_{L\mu}^{\pm}$ and $W_{R\mu}^{\pm}$ is $\sim\frac{v_{LR}^{2}}{v_{R}^{2}}$, it is almost zero. For $v_{R}\sim 10^{8}$  GeV, $M_{W_{R\mu}}$ and $M_{\widetilde{Z}_{\mu}}$ are $\sim 10^{7}$ GeV.

 In the Yukawa sector, the fermion masses and mixings are given by
\begin{alignat}{1}
 & \mathscr{L}_{Fermion Mass}=-(\overline{f_{L}},\overline{\widetilde{f}_{L}})M_{F}(f_{R},\widetilde{f}_{R})^{T}
   -\overline{\nu_{L}}M_{\nu}\nu_{R}-\overline{\widetilde{\nu}_{L}}\frac{M_{\widetilde{\nu}}}{2}\widetilde{\nu}_{R}+h.c.\,,\nonumber\\
 & M_{F}=\left(\begin{array}{cc}-\frac{v_{LR}}{2}e^{\pm i\delta}Y_{Q/L}&-\frac{v_{L}}{\sqrt{2}}Y_{f}\\
   -\frac{v_{R}}{\sqrt{2}}Y_{f}^{\dagger}&M_{\widetilde{f}}\end{array}\right)
   =U_{L}\left(\begin{array}{cc}M^{eff}_{f}&0\\0&M^{eff}_{\widetilde{f}}\end{array}\right)U_{R}^{\dagger}\,,\nonumber\\
 & M^{eff}_{f}\approx-\frac{v_{L}v_{R}}{2}Y_{f}M^{-1}_{\widetilde{f}}Y_{f}^{\dagger}
   =U_{f}\,\mathrm{diag}\left(m_{f_{1}},m_{f_{2}},m_{f_{3}}\right)U_{f}^{\dagger}\,,\hspace{0.5cm}
   M^{eff}_{\widetilde{f}}\approx M_{\widetilde{f}}\,,\nonumber\\
 & M_{\nu}=-\frac{v_{LR}e^{i\delta}}{2}Y_{L}
   =e^{i\delta}U_{\nu}\,\mathrm{diag}\left(m_{\nu_{1}},m_{\nu_{2}},m_{\nu_{3}}\right)U_{\nu}^{\dagger}\,,\hspace{0.5cm}
   M_{\widetilde{\nu}}=-\sqrt{2}\,v_{\phi}Y_{\widetilde{\nu}}\,,
\end{alignat}
 where $f=u,d,e$ and $\widetilde{f}=\widetilde{u},\widetilde{d},\widetilde{e}$. For $(f,\widetilde{f})=(u,\widetilde{u})$, the notation in the first row and first column element of $M_{F}$ takes $e^{+i\delta}Y_{Q}$. For $(f,\widetilde{f})=(d,\widetilde{d})$, the notation is $e^{-i\delta}Y_{Q}$. For $(f,\widetilde{f})=(e,\widetilde{e})$, it is $e^{-i\delta}Y_{L}$. For $Y_{Q/L}\sim10^{-3}$ and $v_{LR}\sim10^{-7}$ GeV, the first row and first column element of $M_{F}$ is indeed very small and ignored, but $M_{\nu}$ can correctly account for the neutrino mass data. For $M_{\widetilde{f}}\approx v_{R}$, $M_{F}$ can be diagonalized as the second equality. The mixing angle between $f_{L}$ and $\widetilde{f}_{L}$ is $\sim \frac{v_{L}Y_{f}}{M_{\widetilde{f}}}\ll 1$, in parallel the mixing angle between $f_{R}$ and $\widetilde{f}_{R}$ is $\sim \frac{v_{R}Y_{f}^{\dagger}}{M_{\widetilde{f}}}\sim 0.1$. In fact, the diagonal structure of $M_{F}$ can also be derived from integrating out the heavier vector-like fermions $\widetilde{f}$ since they have decoupled at the low-energy scale. $M^{eff}_{\widetilde{f}}$ are the effective mass matrices of the vector-like quarks and charged lepton, which approximate to $M_{\widetilde{f}}$, while $M^{eff}_{f}$ are the effective mass matrices of the usual quarks and charged leptons. $M^{eff}_{f}$ are obviously Hermitian matrices, thus they can further be diagonalized by the unitary matrices $U_{f}$. By contrast, there is not any mixing between $\nu$ and $\widetilde{\nu}$ by virtue of the natural characteristics of $\widetilde{\nu}$, so $M_{\nu}$ and $M_{\widetilde{\nu}}$ are directly obtained. Likewise $M_{\nu}$ can be diagonalized by $U_{\nu}$ owing to the $Y_{L}$ Hermiticity. For $v_{\phi}\sim 500$ GeV, in general $M_{\widetilde{\nu}}$ are a few hundred GeVs. It is again stressed that $\nu$ are light Dirac neutrinos in the visible sector, whereas $\widetilde{\nu}$ are heavy Majorana neutrinos in the dark sector, the two species of neutrinos completely differ in nature. It can be seen from (13) that the hierarchical VEVs can lead to such hierarchical mass relations as $M_{\nu}\ll M^{eff}_{f}<M_{\widetilde{\nu}}\ll M_{\widetilde{f}}$. Because $M^{eff}_{f}$ is a quadratic function of $Y_{f}$, some smaller hierarchy of the elements of $Y_{f}$ can naturally lead to some larger hierarchy of the three generation fermion masses. This gives a reasonable explanation for the quark and charged lepton mass hierarchy. The flavor mixing matrix in the quark sector and one in the lepton sector are respectively defined by \cite{27,28}
\ba
 U_{u}^{\dagger}\,U_{d}=U_{CKM}\,,\hspace{0.5cm}
 U_{e}^{\dagger}\,U_{\nu}=U_{PMNS}\,.
\ea
 The mixing angles and $CP$-violating phases in $U_{CKM}$ and $U_{PMNS}$ are parameterized by the standard form in particle data group \cite{1}. Because the origin of $M_{\nu}$ is essentially distinguished from one of $M^{eff}_{f=u,d,e}$\,, the lepton flavor mixing is of course different from the quark one very much. The later numerical results will full demonstrate the interesting features and predictions of the particle masses and mixings.

\vspace{1cm}
 \noindent\textbf{IV. Matter-Antimatter Asymmetry}

\vspace{0.3cm}
 The model can account for the matter-antimatter asymmetry in the universe on the basis of the preceding discussions. As the universe expansion and cooling, the model symmetries are spontaneously broken and reduced step by step. In the evolution process, the matter-antimatter asymmetry will naturally be generated by the following mechanism.

 For most of the inflation models, the reheating temperature is in general $\sim 10^{12-13}$ GeV \cite{29a}. The boson $H_{LR}$ has by nature a very heavy mass about $10^{10}$ GeV, which is below the reheating temperature, therefore the heavy boson $H_{LR}$ can exist in the reheated universe. $H_{LR}$ has an important decay process $H_{LR}\rightarrow L_{L}+\overline{L_{R}}$ in the light of (7) and (8), as shown in Figure 1.
\begin{figure}
 \centering
 \includegraphics[totalheight=4.8cm]{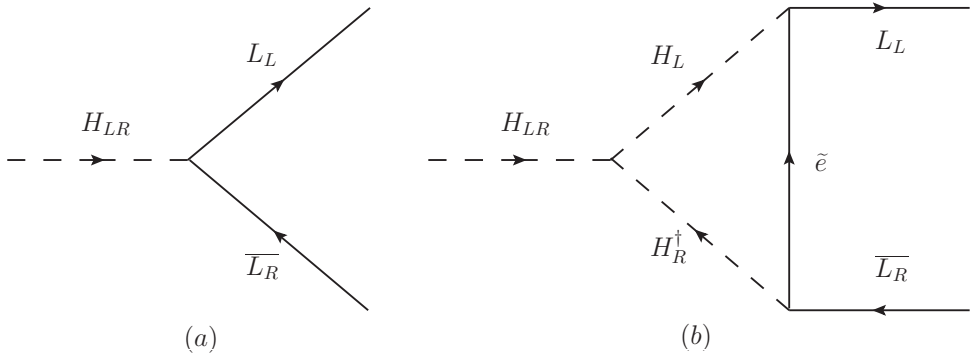}
 \caption{The tree and loop diagrams of the decay $H_{LR}\rightarrow L_{L}+\overline{L_{R}}$, which will lead to the matter-antimatter asymmetry.}
\end{figure}
 Its $CP$ conjugate process, $H_{LR}^{\dagger}\rightarrow\overline{L_{L}}+L_{R}$, has the diagram counterpart as well. The decay process can satisfy two items of Sakharov's three conditions \cite{29b}, namely it is $CP$ violation and out-of-equilibrium but it conserves $B$-$L$ instead of $B$-$L$ violation. The phase $\delta\neq0$  in the scalar potential (8) is explicitly a source of the mirror asymmetry and $CP$ violation. The $CP$-violating phase factor can be transferred into the light neutrino mass $M_{\nu}$ through the spontaneous breakings, see (13). It can surely lead to a $CP$ asymmetry of the decay by the interference between the tree diagram and the loop one. The $CP$ asymmetry is defined and calculated as follow,
\begin{alignat}{1}
 & \varepsilon=\frac{\Gamma(H_{LR}\rightarrow L_{L}+\overline{L_{R}})
   -\Gamma(H_{LR}^{\dagger}\rightarrow\overline{L_{L}}+L_{R})}{\Gamma_{Total}(H_{LR})}
   =\frac{v_{LR}Im[e^{i\delta}\sum\limits_{i=1}^{3}M_{\widetilde{e}_{i}}lnx_{i}(Y_{e}^{\dagger}Y_{L}^{\dagger}Y_{e})_{ii}]}
   {\pi\,v_{L}v_{R}\left(Tr[Y_{L}Y_{L}^{\dagger}+3Y_{Q}Y_{Q}^{\dagger}]+4y^{2}\right)}\,,\nonumber\\
 & \Gamma_{Total}(H_{LR})=\Gamma(H_{LR}\rightarrow L_{L}+\overline{L_{R}})+\Gamma(H_{LR}\rightarrow Q_{L}+\overline{Q_{R}})
   +\Gamma(H_{LR}\rightarrow H_{L}+H_{R}^{\dagger}),\nonumber\\
 & x_{i}=\frac{M_{\widetilde{e}_{i}}}{M_{H_{LR}}}\,,\hspace{0.5cm} y=\frac{\mu_{0}}{M_{H_{LR}}}\,.
\end{alignat}
 For $\mu_{0}\sim10^{4}$ GeV, $M_{H_{LR}}\sim10^{10}$ GeV and $Y_{Q}<Y_{L}\sim10^{-3}$, $H_{LR}\rightarrow L_{L}+\overline{L_{R}}$ is actually the dominant decay process, so $\Gamma_{Total}(H_{LR})\approx\Gamma(H_{LR}\rightarrow L_{L}+\overline{L_{R}})$. For simplicity, the three generation vector-like charged leptons can be taken as mass degeneracy, i.e. $M_{\widetilde{e}_{i}}=M_{\widetilde{e}}$, thus a concise formula of $\varepsilon$ is obtained as
\begin{alignat}{1}
 & \varepsilon=\frac{v_{LR}^{2}M_{\widetilde{e}}^{2}\,lnxIm[Tr(M^{eff}_{e}M_{\nu})]}
   {\pi\,v_{L}^{2}v_{R}^{2}\,Tr[M_{\nu}M_{\nu}^{\dagger}]}\,,\hspace{0.5cm}
   x=\frac{M_{\widetilde{e}}}{M_{H_{LR}}}\,,\nonumber\\
 & Tr(M^{eff}_{e}M_{\nu})=e^{i\delta}Tr[\mathrm{diag}(m_{e_{1}},m_{e_{2}},m_{e_{3}})U_{PMNS}
   \mathrm{diag}(m_{\nu_{1}},\nu_{\nu_{2}},\nu_{\nu_{3}})U_{PMNS}^{\dagger}]\,,\nonumber\\
 & Tr(M_{\nu}M_{\nu}^{\dagger})=\sum\limits_{i=1}^{3}m^{2}_{\nu_{i}}\,,
\end{alignat}
 where the third equation in (16) is given by use of (13) and (14). For $\frac{M_{\widetilde{e}}}{v_{R}}\sim 1$ and the logarithmic dependence on $x$, $v_{LR}$ and $\delta$ are actually the two main parameters in charge of $\varepsilon$. $\varepsilon$ can correctly give a satisfied asymmetry for $v_{LR}\sim 10^{-7}$ GeV and a moderate $\delta$. On the other hand, the out-of-equilibrium condition is that the decay rate is far smaller than the Hubble expansion rate of the universe, namely
\begin{alignat}{1}
   \Gamma(H_{LR}\rightarrow L_{L}+\overline{L_{R}})
 & =\frac{M_{H_{LR}}Tr(Y_{L}Y_{L}^{\dagger})}{16\pi}=\frac{M_{H_{LR}}Tr(M_{\nu}M_{\nu}^{\dagger})}{4\pi v_{LR}^{2}}\nonumber\\
 & \ll H(T=M_{H_{LR}})=\frac{1.66\sqrt{g_{*}}M_{H_{LR}}^{2}}{M_{pl}}\,,
\end{alignat}
 where $M_{pl}=1.22\times10^{19}$ GeV, and $g_{*}$ is the effective number of relativistic degrees of freedom at $T=M_{H_{LR}}$. At this temperature, the non-relativistic particles only include $H_{LR},\widetilde{u},\widetilde{d},\widetilde{e}$, the rest of the model particles are all the relativistic states, thus one can easy figure out $g_{*}=129.25$. For $M_{H_{LR}}\sim10^{10}$ GeV and $v_{LR}\sim10^{-7}$, the condition (17) is indeed satisfied. Besides the two above-mentioned virtues, the decay has another noteworthy characteristic. It violates positive two units of the lepton number for the left-handed lepton doublet $L_{L}$, simultaneously, violates negative two units of the lepton number for the right-handed lepton doublet $L_{R}$, but the total lepton number is conserved. As a result, the decay can respectively lead to a lepton number asymmetry of $L_{L}$ and one of $L_{R}$, which are respectively denoted by $Y_{L_{L}}$ and $Y_{L_{R}}$. $Y_{L_{L}}$ is an asymmetry between the left-handed lepton doublet $L_{L}$ and the right-handed antilepton doublet $\overline{L_{L}}$, and $Y_{L_{R}}$ is a similar meaning. $Y_{L_{L}}$ and $Y_{L_{R}}$ are respectively related to $\varepsilon$ by the first two equations of (18). Because the both are the same size but opposite sign, the total lepton number asymmetry of the universe is still vanishing.

 After the heavy boson $H_{LR}$ decay and decoupling, the model symmetries occur the first step breaking $SU(2)_{R}\otimes U(1)_{Y}\rightarrow U(1)_{Y'}$ at the $v_{R}$ scale, then the gauge symmetries decrease to the SM gauge groups. In this period, the right-handed lepton doublet $L_{R}$ is spontaneously decomposed into the two uncorrelated states $\nu_{R}$ and $e_{R}$, in which $\nu_{R}$ becomes a singlet of the SM. Accordingly, the original asymmetry $Y_{L_{R}}$ is converted into the two separate asymmetries $Y_{\nu_{R}}$ and $Y_{e_{R}}$, where $Y_{\nu_{R}}$ is an asymmetry between the right-handed neutrino and the left-handed antineutrino, and $Y_{e_{R}}$ is a similar meaning. Obviously, $Y_{\nu_{R}}$ and $Y_{e_{R}}$ are the same size and sign, see the second line of (18). Below the temperature of $T=v_{R}$, the reactions as $\nu_{R}+\overline{e_{R}}\rightarrow W_{R}\rightarrow u_{R}+\overline{d_{R}}$ and $\nu_{R}+\overline{\nu_{R}}\rightarrow \widetilde{Z}_{\mu}\rightarrow f+\overline{f}$ are in equilibrium, moreover, the latter reaction can annihilate the symmetry part of the right-handed neutrino but its asymmetry part is left. However, the processes will also be out-of-equilibrium as the universe temperature decreasing on account of $(M_{W_{R}},M_{\widetilde{Z}})\sim 10^{7}$ GeV. As a result, the right-handed neutrinos will completely decouple from the SM particles. The decoupling temperature of $\nu_{R}$ can simply be estimated as $\frac{T_{\nu_{R}}}{T_{\nu_{L}}}\sim(\frac{M_{\widetilde{Z}}}{M_{Z}})^{\frac{4}{3}}$ where $T_{\nu_{L}}$ is the left-handed neutrino decoupling temperature. For $T_{\nu_{L}}\approx1$ MeV, the later numerical calculations will show $T_{\nu_{R}}\sim 10^{4}$ GeV. Because $\nu_{R}$ is relativistic decoupling, $Y_{\nu_{R}}$ is not dependent on the universe temperature in the comoving volume. Therefore, the original $Y_{\nu_{R}}$ does not change before and after $\nu_{R}$ decoupling.

 When the universe temperature decreases to the electroweak breaking scale $v_{L}$, $\nu_{R}$ has been frozen for a long time. The lepton number asymmetry of the SM now consists of $Y_{L_{L}}$ and $Y_{e_{R}}$, namely $Y_{SM}=Y_{L_{L}}+Y_{e_{R}}$, which is obviously non-zero. However, the total lepton number asymmetry of the universe is still vanishing because of $Y_{SM}=-Y_{\nu_{R}}$. At this stage, the sphaleron electroweak transition can smoothly put into effect \cite{30}. It can convert $Y_{SM}$ into the baryon number asymmetry $Y_{B}$, see the last line of (18). By contrast, $Y_{\nu_{R}}$ cannot be converted into $Y_{B}$ because $\nu_{R}$ is not involved in the sphaleron process at all. Thus the original $Y_{\nu_{R}}$ will survive and exists in the present-day universe. Consequently, the universe are eventually generated the two separate matter-antimatter asymmetries $\eta_{B}$ and $\eta_{\nu_{R}}$. In the model, the charged leptons have the effective Yukawa coupling to Higgs as $\overline{L_{L}}H_{L}Y^{eff}_{e}e_{R}$ where $Y^{eff}_{e}=\frac{v_{R}}{\sqrt{2}}Y_{e}M_{\widetilde{e}}^{-1}Y_{e}^{\dagger}$. It can transform the right-handed charged lepton asymmetry into the left-handed one. But there is not an effective Yukawa coupling like this for the light neutrinos. Therefore, the right-handed neutrino asymmetry cannot be transformed into the left-handed one through the so-called mass equilibrium, in other words, it cannot kill the baryon asymmetry. This is not only a source by which the tiny neutrino mass is distinct from the other fermion masses, but also a key point that the baryogenesis works and the right-handed neutrino asymmetry survives. At present day, $\eta_{B}$ has been measured, but $\eta_{\nu_{R}}$ hides itself and eludes the observations. However, $\eta_{B}$ and $\eta_{\nu_{R}}$ are actually a complementary relationship because the both have the same origin. The model predicts $\frac{\eta_{B}}{\eta_{\nu_{R}}}=\frac{28}{51}$. This provides a guide for the future experimental search.

 The above-mentioned procedures and discussions are summarized by the relations as follows,
\begin{alignat}{1}
 & Y_{L_{L}}=\frac{n_{L_{L}}-n_{\overline{L_{L}}}}{s}=\kappa\frac{2\,\varepsilon}{g_{*}}\,,\hspace{0.5cm}
   Y_{L_{R}}=\frac{n_{L_{R}}-n_{\overline{L_{R}}}}{s}=\kappa\frac{-2\,\varepsilon}{g_{*}}\,,\hspace{0.5cm}
   Y_{L_{L}}=-Y_{L_{R}}\,, \nonumber\\
 & Y_{\nu_{R}}=Y_{e_{R}}=\frac{Y_{L_{R}}}{2}\,,\hspace{0.5cm} Y_{SM}=Y_{L_{L}}+Y_{e_{R}}=-Y_{\nu_{R}}\,,\nonumber\\
 & \eta_{B}=\frac{n_{B}-n_{\overline{B}}}{n_{\gamma}}=7.04\,\frac{c_{sp}}{c_{sp}-1}Y_{SM}\,,\hspace{0.3cm}
   \eta_{\nu_{R}}=\frac{n_{\nu_{R}}-n_{\overline{\nu_{R}}}}{n_{\gamma}}=7.04Y_{\nu_{R}}\,,\hspace{0.3cm}
   \frac{\eta_{B}}{\eta_{\nu_{R}}}=\frac{c_{sp}}{1-c_{sp}}\,.
\end{alignat}
 In (18), $\kappa$ is a dilution factor, it can be approximated to $\kappa\approx1$ for the very weak decay. $7.04$ is a ratio of the entropy density $s$ to the photon number density $n_{\gamma}$. $c_{sp}=28/79$ is a coefficient of the sphaleron conversion. Finally, it is again stressed that the asymmetries of matter-antimatter including both baryon and right-handed neutrino take place before the electroweak breaking, in other words, before all the fermion masses are generated, the baryogenesis and the right-handed neutrino asymmetry have been completed. At the later era, therefore the mass equilibration cannot erase the asymmetries of the baryon and right-handed neutrino. Through this mechanism, the universe eventually evolves into the final state with both baryon asymmetry and right-handed neutrino asymmetry from the initial state with the matter-antimatter symmetry. The later numerical results will demonstrate that the baryon asymmetry can successfully be achieved.

\vspace{1cm}
 \noindent\textbf{V. Cold Dark Matter}

\vspace{0.3cm}
 The model can also account for the cold dark matter issue. The model dark sector includes the Majorana neutrino $\widetilde{\nu}$, the pseudo scalar Goldstone boson $\phi_{g}$, and the massive neutral scalar boson $\phi_{s}$ in the light of the model interactions. These particles are neutral singlets under the model gauge groups, in addition, they have the $D$ numbers instead of the $B$-$L$ numbers. By virtue of the characteristics, $\widetilde{\nu}$ and $\phi_{g}$ are actually stable particles in the universe, they respectively become the cold and hot dark matter. $\phi_{s}$ can not only decay into a pair of $\phi_{g}$ or $\widetilde{\nu}$ (if $M_{\phi_{s}}>2M_{\widetilde{\nu}}$), but also decay into a pair of the SM particle. $\phi_{s}$ therefore becomes a messenger between the dark sector and the visible sector.

 In the early universe, The particles $\widetilde{\nu},\phi_{g},\phi_{s}$ are thermal equilibrium in the hot plasma through such reactions as shown Figure 2 (note that $\overline{\widetilde{\nu}}$ identifies with $\widetilde{\nu}$ because of $\widetilde{\nu}$ being Majorana fermion, hereinafter as so).
 \begin{figure}
 \centering
 \includegraphics[totalheight=3.6cm]{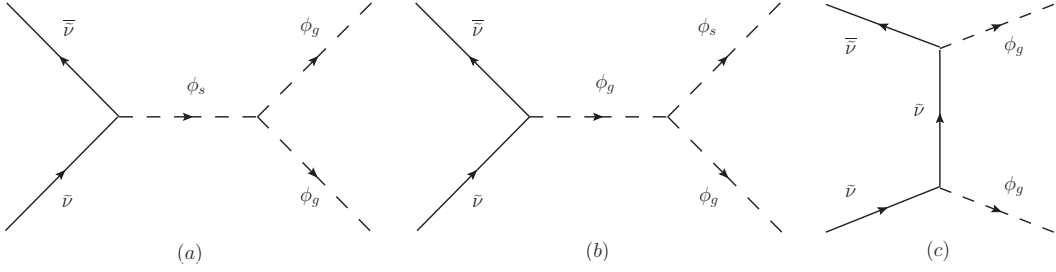}
 \caption{The annihilation ways of a pair of the cold dark matter $\widetilde{\nu}$, in which the diagram (a) is main process.}
\end{figure}
 In addition, a pair of heavier mass $\widetilde{\nu}$ can annihilate into a pair of lighter mass $\widetilde{\nu}$ by the intermediate Goldstone $\phi_{g}$. As the universe temperature decreasing, the reactions will be out-of-equilibrium and $\widetilde{\nu}$ will be frozen. Through careful analysis, the principal annihilation channel of $\widetilde{\nu}$ is actually the diagram (a) in Figure 2, namely a pair of $\widetilde{\nu}$ annihilates into a pair of Goldstone $\phi_{g}$ through the intermediate $\phi_{s}$. Its annihilation cross section is calculated as follows,
\begin{alignat}{1}
 & \sigma_{i}(\widetilde{\nu}_{i}+\overline{\widetilde{\nu}_{i}}\rightarrow \phi_{g}+\phi_{g})
   =\frac{\lambda_{\phi}^{2}\,M_{\widetilde{\nu}_{i}}^{2}}{16\pi[(s-M_{\phi_{s}}^{2})^{2}+(\Gamma_{\phi_{s}}M_{\phi_{s}})^{2}]}
   \sqrt{(1-\frac{4M_{\widetilde{\nu}_{i}}^{2}}{s})}\,,\nonumber\\
 & \Gamma_{\phi_{s}}=\Gamma_{a}(\phi_{s}\rightarrow \phi_{g}+\phi_{g})+\Gamma_{b}(\phi_{s}\rightarrow H_{L}+H_{L}^{\dagger})
   +\sum\limits_{i}\Gamma_{i}(\phi_{s}\rightarrow \widetilde{\nu}_{i}+\overline{\widetilde{\nu}_{i}})\,,\nonumber\\
 & \Gamma_{a}=\frac{\lambda_{\phi}^{2}\,v_{\phi}^{2}}{8\pi M_{\phi_{s}}}\,,\hspace{0.4cm}
   \Gamma_{b}=\frac{\lambda_{2}^{2}\,v_{\phi}^{2}}{32\pi M_{\phi_{s}}}
   \sqrt{1-\frac{4M_{H_{L}}^{2}}{M_{\phi_{s}}^{2}}}\,,\hspace{0.4cm}
   \Gamma_{i}=\frac{M_{\phi_{s}}M^{2}_{\widetilde{\nu}_{i}}}{16\pi v_{\phi}^{2}}
   \left(1-\frac{4M_{\widetilde{\nu}_{i}}^{2}}{M_{\phi_{s}}^{2}}\right)^{\frac{3}{2}},
\end{alignat}
 where $s=4M_{\widetilde{\nu}_{i}}^{2}/(1-v_{i}^{2})$ is the squared center-of-mass energy, and $v_{i}$ is the velocity of $\widetilde{\nu}_{i}$ in the center-of-mass frame. According to the general theory of WIMP \cite{31}, $\widetilde{\nu}$ is non-relativistic decoupling at the freeze temperature $T_{f}\approx \frac{M_{\widetilde{\nu}}}{20}$. The relic abundance of $\widetilde{\nu}$ in the current universe is determined by its annihilation reaction rate \cite{32}. The details are such as
\ba
 \Omega_{D} h^{2}=\sum\limits_{i=1}^{3}\Omega_{i} h^{2}
 \approx \sum\limits_{i=1}^{3}\frac{2.58\times 10^{-10}\,\mathrm{GeV^{-2}}}{\langle\sigma_{i}v_{r}\rangle}\,,
\ea
 where $v_{r}$ is the relative velocity of the two annihilate particles, and the heat average in the denominator can be calculated by $\langle\sigma v_{r}\rangle\approx a+b\langle v_{i}^{2}\rangle=a+b\frac{3T_{f}}{M_{\widetilde{\nu}_{i}}}$. For simplicity, three generations of $\widetilde{\nu}$ can be taken as mass degeneracy, thus $\Omega_{D} h^{2}=3\Omega_{1} h^{2}$. For $M_{\widetilde{\nu}}\sim100$ GeV, $M_{\phi_{s}}\sim300$ GeV, and $\lambda_{\phi}\sim0.1$, (19) can naturally give a weak cross section $\sigma\sim10^{-9}\,\mathrm{GeV^{-2}}$, then (20) will lead to $\Omega_{D} h^{2}\sim0.1$, which meets the current observation data. It should be pointed out that the scenario does not affect BBN since the generation of BBN takes place at the energy scale about $1$ MeV. In a word, $\widetilde{\nu}$ is indeed a good candidate of the cold dark matter.

 I add a brief discussion about two species of the hot dark matter, the light Dirac neutrino and the dark Goldstone boson. In the model, $\nu_{L}$ and $\nu_{R}$ couple together to give rise to the tiny mass. Nevertheless, the destiny of $\nu_{R}$ is very different from one of $\nu_{L}$ due to the mirror symmetry breaking. Firstly, $T_{\nu_{R}}\sim 10^{4}$ GeV is far larger than $T_{\nu_{L}}\sim 1$ MeV. Secondly, the $\nu_{R}$ asymmetry cannot be converted into the baryon asymmetry through the sphaleron process. Thirdly, the relic abundance of $\nu_{R}$ in the current universe is much smaller than one of $\nu_{L}$, a ratio of the both is $\frac{\Omega_{\nu_{R}}}{\Omega_{\nu_{L}}}=\frac{g_{*}(T_{\nu_{L}})}{g_{*}(T_{\nu_{R}})}\approx0.057$ where $g_{*}(T_{\nu_{L}})=6.82$ and $g_{*}(T_{\nu_{R}})=119.25$. At the temperature $T_{\nu_{L}}$ the relativistic states are only $\gamma,\nu_{L},\nu_{R},\phi_{g}$, but at the temperature $T_{\nu_{R}}$ the relativistic states include all of the SM particles and the dark sector particles. For $\Omega_{\nu_{L}}\approx1.7\times10^{-3}$ at the present day, thus one can obtain $\Omega_{\nu_{R}}\approx1\times10^{-4}$, which is about two times the size of the relic abundance of the microwave background photon ($\Omega_{\gamma}\approx5\times10^{-5}$). The Goldstone boson $\phi_{g}$ connection with the SM particles is only a weak coupling of it to the SM Higgs $H_{L}$. At the present era, $\phi_{g}$ has become a species of hot dark matter and only comes into activity in the dark sector. Since it is a massless scalar boson and relativistic decoupling earlier, its effective temperature at the present day is actually lower than the microwave background photon temperature. The current energy density of a relativistic particle is proportional to its effective temperature and degree of freedom, therefore the current abundance of $\phi_{g}$ can be estimated as $\frac{\Omega_{\phi_{g}}}{\Omega_{\gamma}}=\frac{g_{\phi_{g}}T_{\phi_{g}}}{g_{\gamma}T_{\gamma}}<\frac{1}{2}$. In short, $\Omega_{\nu_{R}}+\Omega_{\phi_{g}}<1.25\times10^{-4}$, whereas $\Omega_{\nu_{L}}\approx1.7\times10^{-3}$, the former is an order of magnitude smaller than the latter, so the model is not in conflict with BBN constraints. Finally, all the interesting predictions are a very large challenge for future experimental search.

\vspace{1cm}
 \noindent\textbf{VI. Numerical Results}

\vspace{0.3cm}
 In the section I present the model numerical results. In the light of the foregoing discussions, the model contains a lot of the new parameters besides the SM ones. In principle the SM parameters can be fixed by the experimental data, but the non-SM parameters have yet very large freedoms. The parameters involved in the numerical calculations are now collected together. The gauge sector parameters are the two gauge couplings $g$ and $g_{Y}$. In view of the relevant relations in (12), I can use the mixing angle $tan\widetilde{\theta}$ as a substitute for $g_{Y}$. Furthermore, $g$ and $tan\widetilde{\theta}$ are determined by $e$ and $sin\theta_{W}$ which have precisely been measured by the electroweak physics. The scalar sector parameters include the four coupling coefficients, $\lambda_{\phi},\lambda_{H},\lambda_{2},\lambda_{4}$, and the four VEVs, $v_{\phi},v_{L},v_{R},v_{LR}$, and the mass dimension parameter $\mu_{0}$ and the complex phase $\delta$. Among which, $\lambda_{H}$ and $v_{L}$ are determined by the SM physics and the mass measure of $H_{L}^{0}$ at the LHC. For the non-SM parameters, I only give a set of the typical values instead of the complete analysis for the parameter space. Based on an overall consideration, a set of reasonable and consistent values of the gauge and scalar parameters are chosen as
\begin{alignat}{1}
 & g=0.654\,,\hspace{0.5cm} sin\widetilde{\theta}=0.534\,,\nonumber\\
 & \lambda_{\phi}=0.1\,,\hspace{0.5cm} \lambda_{H}=0.12\,,\hspace{0.5cm}
   \lambda_{2}\leqslant0.01\,,\hspace{0.5cm} \lambda_{4}\leqslant0.01\,,\nonumber\\
 & v_{\phi}=500\:\mathrm{GeV},\hspace{0.4cm} v_{L}=246\:\mathrm{GeV},\hspace{0.4cm}
   v_{R}=1\times10^{8}\:\mathrm{GeV},\hspace{0.4cm} v_{LR}=3\times10^{-7}\:\mathrm{GeV},\nonumber\\
 & \mu_{0}=1\times10^{4}\:\mathrm{GeV},\hspace{0.5cm} \delta=0.117\,\pi\,.
\end{alignat}
 $\lambda_{2}$ and $\lambda_{4}$ are believed to be relatively weaker, especially $\lambda_{2}$ should be bound by the decay $H_{L}^{0}\rightarrow \phi_{g}+\phi_{g}$. However, the condition (10) is satisfied. $\lambda_{\phi}$ and $v_{\phi}$ dominate $M_{\phi_{s}}$ and $M_{\widetilde{\nu}}$ whose limits can be obtained by fitting the relic abundance of the cold dark matter. $v_{LR}$ and $\delta$ are in charge of the baryon asymmetry. Now (21) is substituted into the relevant equations in (11) and (12), the gauge and scalar boson masses are straightforward calculated as follows (in GeV as unit),
\begin{alignat}{1}
 & M_{W_{L}}=80.4\,,\hspace{0.5cm} M_{Z}=91.2\,,\hspace{0.5cm}
   M_{W_{R}}=3.27\times10^{7},\hspace{0.5cm} M_{\widetilde{Z}}=3.87\times10^{7},\nonumber\\
 & M_{\phi_{s}}=223\,,\hspace{0.4cm} M_{H_{L}}=125\,,\hspace{0.4cm} M_{H_{R}}=4.9\times10^{7},\hspace{0.4cm}
   M_{H_{LR}}=2.86\times10^{10}.
\end{alignat}
 Needless to say, these results are very well in accord with the previous discussions. $M_{W_{R}}$, $M_{\widetilde{Z}}$ and $M_{H_{R}}$ are dominated by $v_{R}$. $M_{H_{LR}}$ is affected by $\mu_{0}$. By use of (22), one can obtain $T_{\nu_{R}}\approx3.2\times10^{4}$ GeV.

 The Yukawa sector contains a great deal of the flavor parameters. However, I can choose such flavor basis as all of $M_{\widetilde{u},\widetilde{d},\widetilde{e}}$ and $Y_{\widetilde{\nu},L,Q}$ being diagonal. In addition, $Y_{\widetilde{\nu}}$ and $Y_{L}$ can be replaced by $M_{\widetilde{\nu}}$ and $M_{\nu}$ in the light of (13). The couplings $Y_{u,d,e}$ implicate the flavor structures, and determine the masses and mixings of the quarks and charged leptons. The calculations of the baryon asymmetry involve in the mass and mixing parameters of the charged leptons and light neutrinos, which have excellently been measured except the two undetermined parameters \cite{1}. One parameter can be chosen as $m_{\nu_{2}}$ since the two mass-squared differences $\triangle m^{2}_{21}$ and $\triangle m^{2}_{32}$ are known. The other is the $CP$-violating phase $\delta^{l}$ in the lepton flavor mixing, which has no any information from the experiments by now. For simplicity, $M_{\widetilde{e}}$ and $M_{\widetilde{\nu}}$ are taken as two constant unit matrices, i.e. mass degeneracy. Furthermore, $M_{\widetilde{e}}$ is fixed to the same value as $v_{R}$ for the theoretical consistency. $M_{\widetilde{\nu}}$ is determined by fitting the relic abundance of the cold dark matter. The values of the Yukawa sector parameters are as follows,
\begin{alignat}{1}
 & m_{e}=0.511\:\mathrm{MeV},\hspace{0.5cm} m_{\mu}=105.7\:\mathrm{MeV},\hspace{0.5cm}
   m_{\tau}=1777\:\mathrm{MeV},\nonumber\\
 & m_{\nu_{2}}=0.01\:\mathrm{eV},\hspace{0.5cm}
   \triangle m^{2}_{21}=7.6\times10^{-5}\:\mathrm{eV^{2}},\hspace{0.5cm}
   \triangle m^{2}_{32}=2.35\times10^{-3}\:\mathrm{eV^{2}},\nonumber\\
 & sin\theta_{12}=0.558\,,\hspace{0.5cm} sin\theta_{23}=0.7\,,\hspace{0.5cm}
   sin\theta_{13}=0.158\,,\hspace{0.5cm} \delta^{\,l}=0\,,\nonumber\\
 & M_{\widetilde{e}}=1\times10^{8}\:\mathrm{GeV},\hspace{0.4cm}
   M_{\widetilde{\nu}}=100.7\:\mbox{or}\:124\:\mathrm{GeV}.
\end{alignat}
 By use of (16)-(20), now the baryon asymmetry and the relic abundance of the cold dark matter $\widetilde{\nu}$ are calculated as
\ba
 \frac{\Gamma(H_{LR}\rightarrow L_{L}+\overline{L_{R}})}{H(T=M_{H_{LR}})}=0.052\,,\hspace{0.5cm}
 \eta_{B}=6.15\times10^{-10}\,,\hspace{0.5cm} \Omega_{D}h^{2}=0.112\,,
\ea
 where the ratio of the decay width of $H_{LR}\rightarrow L_{L}+\overline{L_{R}}$ to the Hubble expansion rate is also given. It is very clear that the decay process is indeed out-of-equilibrium. The values of $\eta_{B}$ and $\Omega_{D}h^{2}$ are precisely in agreement with the current data of the baryon asymmetry and cold dark matter \cite{33}. $M_{\widetilde{\nu}}$ has double value solution in (23). The smaller value corresponds to $2M_{\widetilde{\nu}}<M_{\phi_{s}}$, while the larger value corresponds to $2M_{\widetilde{\nu}}>M_{\phi_{s}}$.

\begin{figure}
 \centering
 \includegraphics[totalheight=8.8cm]{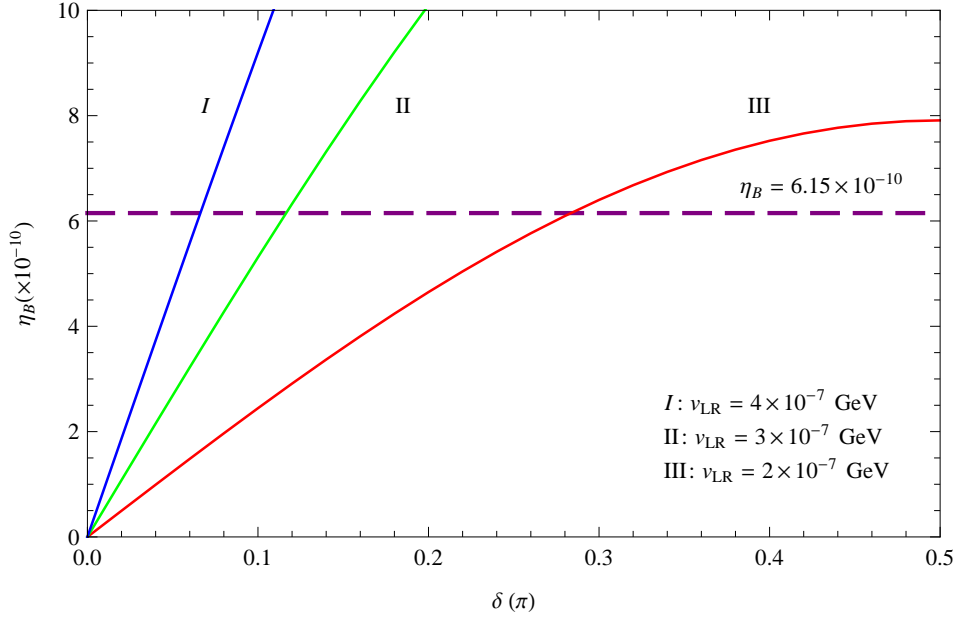}
 \caption{The graphs of the baryon asymmetry subjecting to the phase $\delta$ for $v_{LR}=(2\times10^{-7},3\times10^{-7},4\times10^{-7})$ GeV, while the other parameters are fixed by (21) and (23). The curve II corresponds to the case in the context.}
\end{figure}
 Figure 3 draws $\eta_{B}$ subjecting to $\delta$ for the three values of $v_{LR}=(2\times10^{-7},3\times10^{-7},4\times10^{-7})$ GeV, while the other parameters are fixed by (21) and (23). The intersection of the curve II and the horizontal baseline of $\eta_{B}=6.15\times 10^{-10}$ exactly corresponds to the values of $v_{LR}$ and $\delta$ in (21). It can be seen from Figure 3 that $\eta_{B}$ increases with increasing $v_{LR}$ when $\delta$ is fixed, in other words, $\delta$ decreases with increasing $v_{LR}$ along the $\eta_{B}$ baseline. Thus, a reasonable region of $v_{LR}$ should be $\sim (2\times10^{-7}-4\times10^{-7})$ GeV for the moderate $\delta$. In particular, it should be pointed out that the leptonic $CP$-violating phase $\delta^{l}$ varying has nearly no effect on $\eta_{B}$, in other words, in essence the baryon asymmetry has nothing to do with the lepton $CP$ violation. Therefore, I can draw a conclusion that the matter-antimatter asymmetry completely originates from the mirror symmetry breaking and $CP$ violation in the scalar sector.

\begin{figure}
 \centering
 \includegraphics[totalheight=8.8cm]{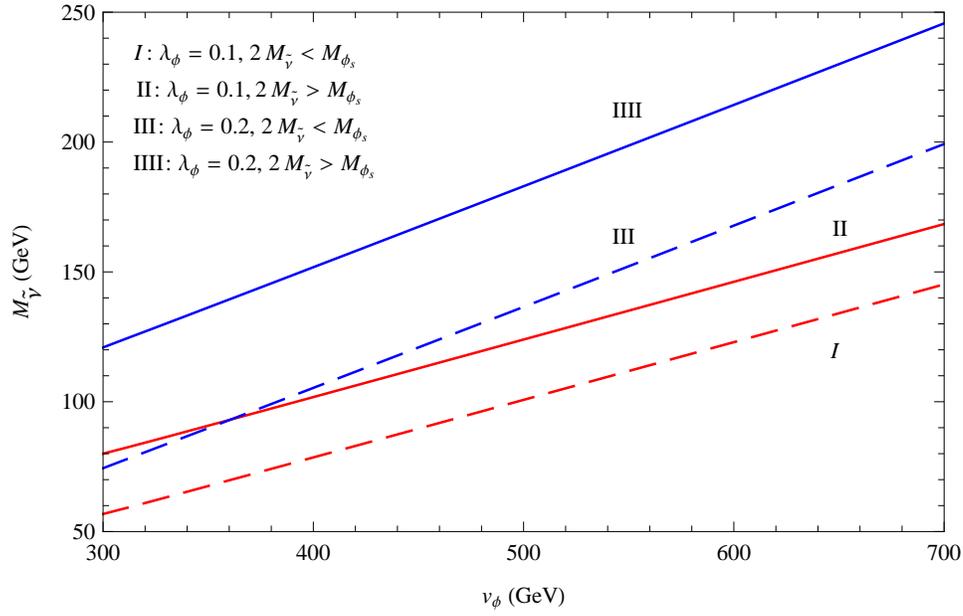}
 \caption{The graphs of $v_{\phi}$ versus $M_{\widetilde{\nu}}$ satisfying $\Omega_{D}h^{2}=0.112$ for $\lambda_{\phi}=(0.1,0.2)$, while the other parameters are fixed by (21) and (23). The dashed and solid lines correspond to the case of $2M_{\widetilde{\nu}}<M_{\phi_{s}}$ and $2M_{\widetilde{\nu}}>M_{\phi_{s}}$, respectively.}
\end{figure}
 Figure 4 is the graphs of $v_{\phi}$ versus $M_{\widetilde{\nu}}$ satisfying $\Omega_{D}h^{2}=0.112$, in which $\lambda_{\phi}$ is taken as $0.1$ and $0.2$, while the other parameters are fixed by (21) and (23). The dashed and solid lines correspond to the case of $2M_{\widetilde{\nu}}<M_{\phi_{s}}$ and $2M_{\widetilde{\nu}}>M_{\phi_{s}}$, respectively. At $v_{\phi}=500$ GeV, the two points on the red dashed and solid lines exactly corresponds to the two values of $M_{\widetilde{\nu}}$ in (23). $M_{\widetilde{\nu}}$ increases with increasing $\lambda_{\phi}$ when $v_{\phi}$ is fixed. Evidently, a smaller $v_{\phi}$ cannot meet the model theoretical requirements and the experimental limits, whereas a larger $v_{\phi}$ can lead to a larger $M_{\widetilde{\nu}}$, this is also unacceptable. Based on an overall consideration, a reasonable area of $v_{\phi}$ and $M_{\widetilde{\nu}}$ should be $v_{\phi}\sim(300-800)$ GeV and $M_{\widetilde{\nu}}\sim(50-300)$ GeV. Because $\widetilde{\nu}$ has no any direct interactions with the SM particles, it will be very difficult for searching it in future experiments. However, $v_{\phi}$ is able to be determined by finding the boson $\phi_{s}$ because it has a weak coupling to the SM Higgs $H_{L}$.

 A whole mass spectrum of all kind of the model particles is now summarized as follows,
\begin{alignat}{1}
 & M_{A_{\mu}}=M_{\phi_{g}}=0\,,\hspace{0.5cm} M_{\nu}\sim 0.01\:\mathrm{eV},\nonumber\\
 & M_{SM particles}\sim(0.001-100)\:\mathrm{GeV},\nonumber\\
 & M_{\phi_{s}}\sim(130-350)\:\mathrm{GeV},\hspace{0.5cm} M_{\widetilde{\nu}}\sim(50-300)\:\mathrm{GeV},\nonumber\\
 & (M_{W_{R}},M_{\widetilde{Z}},M_{H_{R}},M_{\widetilde{f}})\sim10^{7-8}\:\mathrm{GeV},\hspace{0.5cm}
   M_{H_{LR}}\sim10^{10}\:\mathrm{GeV}.
\end{alignat}
 All the numerical results clearly demonstrate the main ideas of the model. The model can not only completely accommodate all of the measured data of the SM and neutrino physics, but also accurately reproduce the observed data of the baryon asymmetry and cold dark matter. All the results are naturally produced without any fine tuning.

 In the end, I give some methods how to search these non-SM particles $\nu_{R},\phi_{s},\phi_{g},\widetilde{\nu}$ at the LHC. On the basis of the model interactions, Figure 5 draws some feasible production and decay processes.
\begin{figure}
 \centering
 \includegraphics[totalheight=4.8cm]{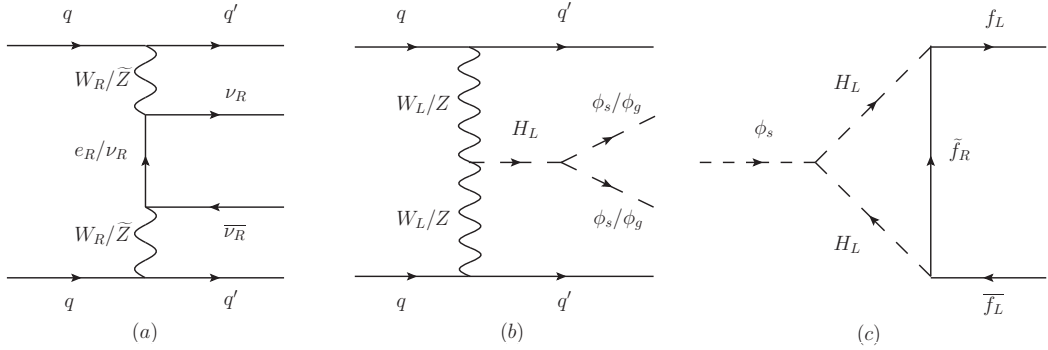}
 \caption{(a) the pair production of $\nu_{R}$ by the proton-proton collisions, (b) the pair production of $\phi_{s}$ or $\phi_{g}$, (c) the decay of $\phi_{s}$ into a pair of the quark or charged lepton.}
\end{figure}
 The diagram (a) illustrates the pair production of $\nu_{R}$ by the proton-proton collisions, of course, this process cross section is so tiny that detection to $\nu_{R}$ will be very difficult. The diagram (b) can produce a pair of $\phi_{s}$ or $\phi_{g}$, however, its cross section is also tiny because of the weak coupling $\lambda_{2}$. $\phi_{s}$ principally decays into a pair of $\phi_{g}$ or $\widetilde{\nu}$ (if $2M_{\widetilde{\nu}}<M_{\phi_{s}}$), and it can also decay into a pair of the quark or charged lepton, as shown the diagram (c). Therefore, the loss of energy in the $\phi_{s}$ decay should be regarded as a definitive signal of the dark matter neutrino $\widetilde{\nu}$ and Goldstone $\phi_{g}$. Although all of the searches are very challenging for the future experiments, the model is feasible and promising to be tested in the near future.

\vspace{1cm}
 \noindent\textbf{VII. Conclusions}

\vspace{0.3cm}
 In the paper, I suggest a simple and feasible particle model. The model extends the SM to the left-right mirror symmetric theory with the global $U(1)_{B-L}\times U(1)_{D}$ symmetries. The heavy scalar $H_{LR}$ in the model develops a tiny VEV after the gauge symmetry breakings, by which the tiny mass of the light neutrino is generated. The decay of $H_{LR}\rightarrow L_{L}+\overline{L_{R}}$ is $CP$ violation and out-of-equilibrium. The $CP$-violating source is in the scalar potential. Through the mirror symmetry breaking and sphaleron process, this eventually leads to both the baryon asymmetry and the light right-handed neutrino asymmetry. In addition, the $U(1)_{D}$ symmetry breaking causes that the Majorana neutrino $\widetilde{\nu}$ and Goldstone boson $\phi_{g}$ become the cold and hot dark matter, respectively. The model can not only naturally accommodate the SM and neutrino physics, but also elegantly account for the matter-antimatter asymmetry and cold dark matter. The four things are closely interrelated and completely integrated in this theory. In particular, the model gives a number of interesting predictions, for instance, the light right-handed neutrino asymmetry is the same size as the baryon asymmetry, its relic abundance is the same size as one of the microwave background photon, the cold dark matter neutrino mass is $\sim(50-300)$ GeV, the neutral scalar boson occurs around (130-350) GeV, and so on. However, the reason and mechanism of the mirror symmetry breaking are yet unknown in the model. They maybe have something to do with some underlying physics, e.g. string theory. This is worthy of further research. Finally, it is feasible and promising to test the model in future experiments. Some of the non-SM particles will possibly be discovered in the future. Undoubtedly, any progress toward these goals will promote our understandings to the mysteries of the universe.

\vspace{1cm}
 \noindent\textbf{Acknowledgments}

\vspace{0.3cm}
 I would like to thank my wife for her large helps. This research is supported by chinese universities scientific fund.

\end{document}